# 3-way equal filtering power divider using compact folded-arms square open-Loop resonator

Augustine O. Nwajana
*Dept. of Electrical and Electronic*
*School of Engineering*
*University of Greenwich*
London, UK
a.o.nwajana@greenwich.ac.uk

Rose Paul
*Dept. of Electrical and Electronic*
*School of Engineering*
*University of Greenwich*
London, UK
rp1009i@greenwich.ac.uk

*Abstract*—Microstrip three-way (that is, 4.8 dB) integrated filtering power divider (FPD) is presented in this paper. The proposed FPD evenly distributes an input power signal into three equal output signals. The design incorporates balanced signal power division, and filtering technology for the removal of unwanted frequency elements and aimed at enhancing signal quality and efficiency in the radiofrequency (RF) front-end of communication systems. Microstrip folded-arms square open-loop resonator (FASOLR) is employed in the design implementation to achieve compact size. The proposed FPD features a 2.6 GHz centre frequency, with a 0.03 fractional bandwidth. The implementation is carried out on Rogers RT/Duroid 6010LM substrate with a dielectric constant of 10.7, a thickness of 1.27 mm and a loss tangent of 0.0023. The good agreement between the theoretical and practical results verifies the effectiveness of the FPD in delivering equal power outputs at the three output ports, and at the same time filtering out unwanted frequencies. The practical results of the prototype FPD indicate a good return loss of better than 15.5 dB and an insertion loss of better than 4.77+0.34 dB. The design prototype achieved compact size of 0.31 $\lambda g$ x 0.18 $\lambda g$. $\lambda g$ is the guided wavelength for the microstrip line impedance at the centre frequency of the 3-way equal filtering power divider.

I. INTRODUCTION

In the active environment of communication systems, the challenge of effective power division and allocation, and correct signal filtering has increased. Wilkinson Power Divider (WPD) is the traditional power division approach using quarter-wavelength transmission lines [1]. This type of power divider is popularly used in communication systems due to its good electrical isolation and simple structure [2]. The problem with WPD is that its operation in a communication system requires the use of an externally connected bandpass filter as shown in Fig. 1(a). This technique of using a transmission line to externally link the WPD with a filter will not only increase the physical size but will also lead to higher loss coefficient in the system [3]. In contrast to the traditional approach involving a bandpass filter, a matching transmission line, and a Wilkinson power divider as separate components, the proposed integrated filtering power divider (FPD) consolidates all three components into a single device [4]–[6] as shown in Fig. 1(b). The integrated FPD means reduce complexity, physical size, and loses.

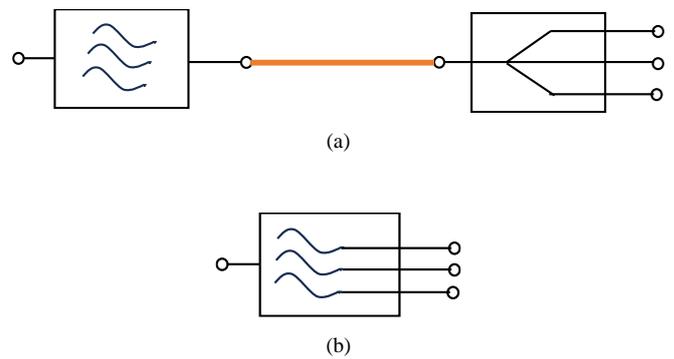

Fig. 1. Three-way power divider (a) Conventional cascaded bandpass filter, transmission line, and three-way Wilkinson power divider, (b) proposed integrated three-way filtering power divider.

Numerous Wilkinson power dividers have been proposed using various technologies. One such technology for crafting a 3-way wideband FPD replaces traditional WPD's quarter-wavelength (λ/4) transformers with multiple coupled lines [5], [7]–[9]. The design seeks to mitigate concerns related to circuit dimensions and insertion losses, offering an asymmetric 3-way equal filtering power divider with high selectivity by employing two dissimilar quarter-wavelength transformers, one pair of short-ended parallel-coupled microstrip lines, one open-ended parallel-coupled line, and one 3-line parallel-coupled line to achieve equal energy subdivision. However, there is some variation between electromagnetic (EM) and circuit responses due to not considering open ends, and cross junctions in the ideal circuit [5]. A 3-line coupled configuration and λ/4 open terminals are employed in the modelling of a compact three-way wideband FPD featuring a 1:1:2 power ratio. It integrates parallel λ/2 open-circuit transmission lines for effective filtering. The design's flexibility is highlighted by independent control of centre frequency and bandwidth for each passband through resonator parameters [7]. Whereas in [8], a compact 3-way power divider with wideband bandpass filtering was reported and it utilizes short-circuited-stub (SCS) loaded unequal power divider, unequal-width 3-coupled lines, and isolation resistors. Nonreciprocal filtering power dividers are introduced in [9], which are compact devices that combine power division, combination, and bandpass filtering with magnet-less non-reciprocity using time-modulated varactors.

A microstrip prototype with a Wilkinson power divider and modulated quarter-wavelength coupled line filters exhibits electrical reconfigurability, high rejection beyond the passband, and strong port isolation. The low quality-factor (Q-factor) typically exhibited by microstrip printed components causes high losses in the device. Improving the Q-factor of the varactors or using alternative resonator structures could enhance the overall performance of the device.

Substrate integrated waveguide (SIW) [10]–[13] is another technology recently deployed in the implementation of filtered power dividers. This technology is compatible with single-layer PCB technology as reported in [10], [11]. A single-layer 3-way SIW FPD that achieved equal power distribution by using degenerated modes in rectangular SIW cavities was proposed in [12]. Employing waveguide theory, microstrip inset feedlines, and open-circuited stubs for impedance matching, the design achieved equal power division with good performances. However, the presence of a spurious mode at the operating frequency, despite efforts to suppress it, may have an impact on the stopband rejection and overall signal integrity. A three-way FPD using multilayer SIW technology was reported in [13]. This achieved both unequal and equal power division by integrating slotline-to-microstrip transition arrangements with TE101 mode resonant cavities. The prototypes demonstrated effective performance with wide bandwidth, low insertion loss, and high in-band isolation, showcasing potential applications in advanced wireless communication systems. However, the complexity of the proposed design may limit its practical implementation in certain applications.

There have been proposals for Power dividers incorporating Composite right and left-handed (CRLH) structures [14], [15]. A compact dual-band three-way metamaterial power divider with an equal output phase was proposed in [14]. This employs fully printed CRLH structures for 900 MHz and 2.4 GHz bands, incorporating 2:1 unequal and 1:1 equal power dividers. The design initially introduced phase errors at three outputs, which were later corrected using a CRLH phase-compensating line. [15] reported a compact 3-way Bagley polygon power divider with a built-in bandpass response. It used innovative CRLH transmission lines for resonators and zero-degree transmission lines to shrink the circuit size while preserving isolation performance. The design ensures in-phase output ports and a broad upper stopband.

Various other techniques have been employed in designing power dividers to showcase enhanced performance and compactness [16]–[20]. Miniaturization of power dividers can be achieved using defected ground structures (DGS). Split ring resonators (SRR) are used as DGS in a three-way power divider to reduce its size. Unlike traditional methods involving loaded lines and repetitive structures, strategically placing defects achieves similar performance in a more compact design. This technique suggests potential applications of DGS in different microwave devices including diplexers and couplers and diplexers [16]. Two symmetric F-shaped DGS are employed in a 3-way planar WPD divider which demonstrates strong overall ultra-wideband performance [17]. A three-way power splitter that uses a multimode patch resonator is proposed in [18]. This utilizes the unique characteristics of the equilateral triangle patch resonator modes for power division. To achieve the desired two-order filtering response, the design incorporates metal vias and slots to adjust the resonant frequencies. In [19], a unique three-way power divider featuring bandpass responses, employing folded net-type resonators to achieve compact size while serving twin functions of filtering and power division was proposed. The symmetric structure ensures equal power distribution with minimal magnitude imbalance. The design of a 2-way and a 3-way FPDs using regular triangle patch resonator (RTPR) was proposed in [20]. The selection of resonator modes is based on their electric-field distributions, showcasing the RTPR's flexibility. The design incorporates an in-phase multiway excitation scheme and strategically placed shorting posts for effective harmonic rejection, resulting in significantly extended stopbands.

This paper proposes a compact 3-way filtering power divider (or combiner) that delivers equal amount of power to each of its output ports, while at the same time attenuating unwanted frequency components. Hence, the proposed integrated FPD acts as a three-in-one device since it consolidates three components (a bandpass filter, a matching transmission line, and a Wilkinson power divider) into a single device as shown in Fig. 1. The proposed device removes the complexity of having three separate devices in a communication system by replacing all three with a single device, leading to reduced footprint in the system. The design also leads to reduced losses as the conventional tuning required in the matching transmission line of Fig. 1(a) is absent as seen in Fig. 1(b). To achieve further miniaturisation, the proposed design implementation employs the compact microstrip (MS) folded-arms square open-loop resonator (FASOLR) reported in [21]. The FASOLR has been found to have a small side length of 1/8 the length of the conventional quarter-wavelength resonator employed in the realisation of the traditional WPD reported in [1].

## II. THEORETICAL CIRCUIT CONFIGURATION

The circuit configuration for the proposed integrated FPD is achieved by first designing three identical bandpass filters using the standard normalised Chebyshev lowpass prototype filter element-values of g0 = g4 = 1.0, g1 = g3 = 0.8516, and g2 = 1.1032. The design centre frequency is chosen as 2.6 GHz, with a fractional bandwidth (FBW) of 0.03, and the input/output characteristic impedance (Z0) of 50 Ohms. The identical filter circuits were designed based on the technique reported in [22], and then coupled together to achieve the integrated filtering power divider circuit arrangement shown in Fig. 2.

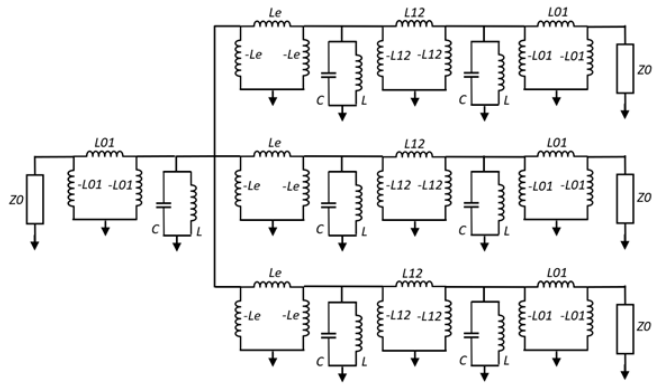

Fig. 2. Circuit arrangement for the proposed filtering power divider (Z0 = 50 Ω, L01 = 3.0607 nH, L12 = 3.4836 nH, Le = 6.0337 nH).

## III. LAYOUT AND PRACTICAL ARRANGEMENT

The equal FPD design and simulation were carried out using microstrip technology and the FASOLR was used to implement the layout. The width and guided wavelength for designing the FASOLR are determined from [21]. PathWave Advanced Design System (ADS) EM Simulation Software was used to create and simulate the equal FPD layout model, with the resonators coupling arrangement shown in Fig. 3 representing the design foundation. The layout was created on Rogers RT/Duroid 6010LM substrate with a dielectric constant of 10.7, a thickness of 1.27 mm, and a loss tangent of 0.0023. The FASOLR is designed to resonate at the circuit model specified centre frequency of 2.6 GHz.

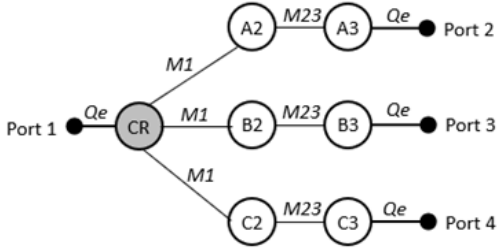

Fig. 3. Coupling arrangement for the proposed equal filtering power divider.

$$M_{23} = \frac{FBW}{\sqrt{g_1 g_2}} = 0.031 \quad (1)$$

$$M_1 = \frac{M_{23}}{\sqrt{3}} = 0.0179 \quad (2)$$

$$Q_e = \frac{g_0 g_1}{FBW} = 28.387 \quad (3)$$

As shown in the coupling scheme of Figure 3, it is necessary to determine two coupling coefficient values M23 and M1 using (1) and (2), respectively. M23 is the coupling coefficient between each resonator pair in the three identical bandpass filters. M1, on the other hand, is the coupling coefficient between the common resonator (CR) and each filter branch. M1 ensures the equitable distribution of the input power entering port 1 amongst the three output ports. It is important to note that CR contributes three poles to the FPD responses as it replaces one resonator from each of the three identical bandpass filters. This means that CR contributes the reduced physical footprint of the proposed FPD device. The external quality factor (Qe) is the coupling between the input port (that is, port 1) and CR. It is also the coupling between the last resonator of each bandpass filter and its corresponding output port (that is, ports 2, 3, or 4) indicated in Fig. 3. The coupling value for Qe is determined using (3) which is derived from [23]. The physical dimensions of the equal FPD microstrip layout are shown in Fig. 4(a), with the pictorial image of the fabricated device shown in Fig. 4(b).

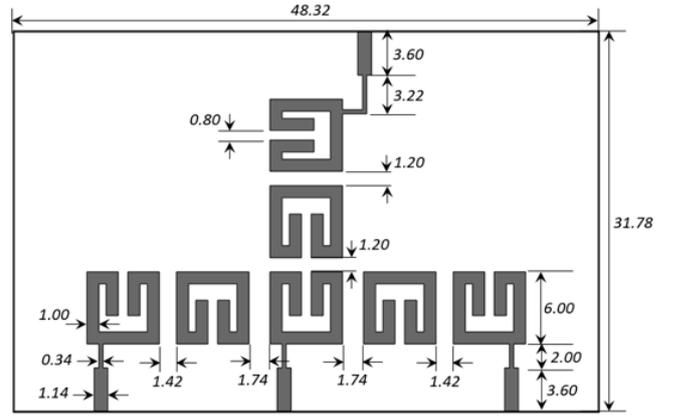

(a)

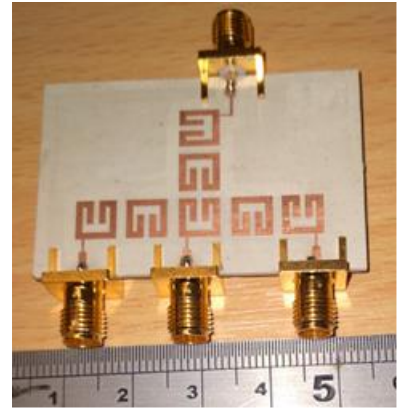

(b)

Fig. 4. Microstrip layout of the three-way equal filtering power divider (a) Dimensions in mm, (b) Pictorial image of the fabricated device.

## IV. RESULTS ANALYSIS AND DISCUSSION

This section captures and discusses both the theoretical and practical results for the proposed equal filtering power divider. The results are jointly presented in Fig. 5 for ease of analysis and comparison. The proposed equal FPD performances captured in Fig. 5 shows good agreement between the theoretical and practical responses. The solid line plots in the Fig. 5 represent the theoretical results while the dashed line plots denote the practical results. Looking at the differential mode responses of Fig. 5(a), a practical minimum return loss of 15.5 dB is achieved compared to the 20dB return loss of the theoretical/ideal results. The achieved practical minimum insertion loss is recorded as 4.77+0.34 dB, where the 4.77 dB part is the standard theoretical inherent characteristics return loss of any three-way equal filtering power divider device as reported in [3]. The achieved variation in return losses of 0.34 dB between the theoretical and the practical equal FPD results is very good when compared to the state-of-the-art as captured in Table 1. The single-ended output ports responses shown in Fig. 5(b) also confirms the good agreement between the theoretical and practical results of the proposed equal filtering power divider.

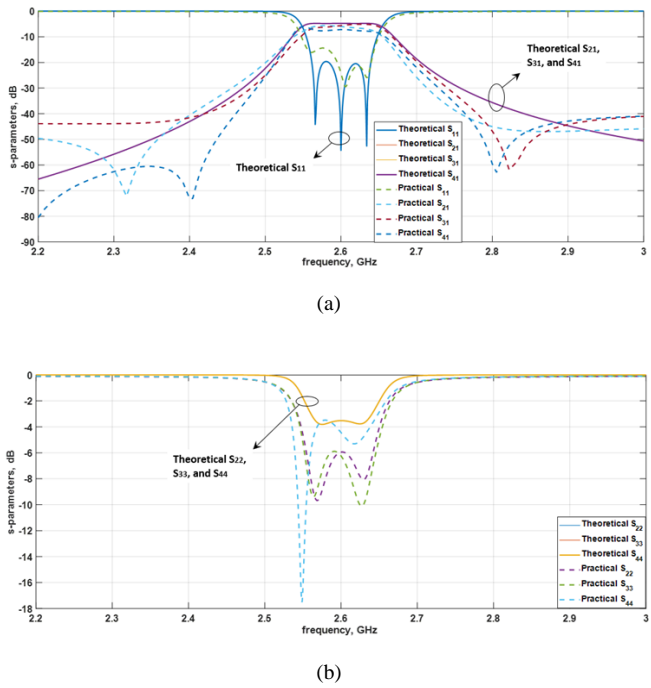

Fig. 5. Theoretical and practical results comparison (a) Differential mode results, (b) Single ended output ports results.

TABLE I. PERFORMANCE COMPARISON

| Ref. | $f_0$ (GHz) | Filter order | Size $(\lambda_g)^2$ | TL[a] | RL[b] (dB) | IL[c] (dB) |
|---|---|---|---|---|---|---|
| [5] | 1.36 | 3 | 0.1853 | MS | 19.5 | 0.40 |
| [7] | 3.50 | 3 | 0.1440 | MS | 15.0 | 2.18 |
| [8] | 1.50 | 4 | 0.0825 | MS | 14.2 | 0.80 |
| [12] | 6.45 | 2 | 0.1593 | SIW | 13.6 | 1.10 |
| [13] | 11.80 | 2 | 3.6666 | SIW | 16.3 | 0.80 |
| [15] | 3.50 | 2 | 1.2789 | MS | 14.5 | 1.43 |
| This work | 2.60 | 3 | 0.0558 | MS | 15.5 | 0.34 |

[a] transmission line, [b] return loss, [c] insertion loss.

## V. CONCLUSION

A balanced three-way filtering power divider that receives an input power and delivers an equal amount of output to its three output ports has been developed and realized using microstrip technology. The design implementation is based on the compact folded-arms square open-loop resonators (FASOLR) for device miniaturization. The proposed filtering power divider operates at a 2.6 GHz centre frequency, with a 3% fractional bandwidth. The realization and fabrication are carried out on Rogers RT/Duroid 6010LM substrate with a dielectric constant of 10.7, a thickness of 1.27 mm and a loss tangent of 0.0023. The theoretical and practical results of the proposed filtering power divider indicate good agreement with a return loss of better than 15.5 dB and an insertion loss of better than 4.77+0.34 dB. The fabricated device achieved a compact size of 0.31 $\lambda_g$ x 0.18 $\lambda_g$. The compact design is well-suited for modern multifunctional communication systems due to its low insertion and return losses, sharp selectivity, and integrated filtering capability.